# The different neighbours around Type-1 and Type-2 active galactic nuclei

*Authors: Beatriz Villarroel & Andreas J. Korn*

Department of Physics and Astronomy, Uppsala University, SE-751 20 Uppsala, Sweden

– 2 –

## ABSTRACT

One of the most intriguing open issues in galaxy evolution is the structure and evolution of active galactic nuclei (AGN) that emit intense light believed to come from an accretion disk near a super-massive black hole (1; 2).

To understand the zoo of different AGN classes, it has been suggested that all AGN are the same type of object viewed from different angles (3). This model – called AGN unification – has been successful in predicting e.g. the existence of hidden broad optical lines in the spectrum of many narrow-line AGN.

But this model is not unchallenged(4) and it is an open problem whether more than viewing angle separates the so-called Type-1 and Type-2 AGN.

Here we report the first large-scale study that finds strong differences in the galaxy neighbours to Type-1 and Type-2 AGN with data from the Sloan Digital Sky Survey (SDSS) (5) Data Release 7 (DR7) (6) and Galaxy Zoo (7; 8).

We find strong differences in the colour and AGN activity of the neighbours to Type-1 and Type-2 AGN and in how the fraction of AGN residing in spiral hosts changes depending on the presence of a neighbour or not.

These findings suggest that an evolutionary link between the two major AGN types might exist.

## 1. Letter

Much of our understanding of the AGN structure relies on the AGN unification model. The model has been successful in predicting e.g. the existence of hidden broad optical lines in the spectrum of many narrow-line AGN, as well as ionization cones and the isotropy of the [O III]5007 line emission. It was developed after the detection of a hidden broad-line region (BLR) in the spectrum of a Seyfert 2 galaxy when observed in polarized light (9). This indicated that the light from the accretion disk passes through optically thick material on the way to the observer. In the extreme simplification of Antonucci's model (3), the accretion disk and BLR can be hidden from the observer's view when embedded in a donut-shaped dust torus. The Type-1 AGN (broad-line) are viewed face-on to the opening of the torus, while in Type-2 AGN (narrow-line) one faces the obscuring part of it. In a more realistic scenario, the covering factors of the torus differ for each individual AGN (10). Also tori consisting of many individual dust clouds having different covering factors have been considered (11). These differences in the observables prevent us from directly comparing the intrinsic properties of the two types of AGN due to mass-to-luminosity biases.

An unresolved issue has been the subject of much controversy: if the viewing angle is all that separates objects with otherwise identical AGN properties, why do only 50% of the Type-2 AGN reveal a hidden BLR (4)? Some common explanations have so far been extremely low accretion rates (12; 13) and extreme obscuration (14). So are Type-1 and Type-2 AGN truly representing the same kind of object?

The main idea of our hypothesis is that if Type-1 and Type-2 AGN are intrinsically the same objects only viewed from different angles, their neighbours should, in a statistical sense, not differ systematically. On top of this, the AGN should interact in similar ways with them.

We construct parent samples of broad-line (Type-1) and narrow-line (Type-2) AGN at redshifts $0.03 < z < 0.2$ using optical emission line classification methods (Balmer line width and Kauffmann criteria (15)) based on data from SDSS DR7. This gives us in total 11334 Type-1 and 53416 Type-2 AGN. It is important to note, that the AGN Unification does not unify radio-loud and radio-quiet AGN. However, the vast majority of AGN in spiral hosts are radio-quiet and are at these luminosities classified as Seyfert galaxies. We therefore also make use of the morphological classifications of the AGN host galaxies from Galaxy Zoo where the hosts are classified either as "Spiral", "Elliptical" or "Uncertain" based on voting percentages (at least 80%). The "Uncertain" category includes those objects classified as neither "Spiral" nor "Elliptical". We verify that the samples in volume-limited subsample cuts have qualitatively similar distributions in redshift, absolute magnitude and luminosity $L(\text{[O\,{\sc iii}]}5007)$, see Fig. S1-S3 (Supp. Inf.). Neighbours with spectroscopic redshifts are selected with the redshift difference cut $|\Delta z| < 0.012$ and within a projected distance of 350 kpc, yielding 1658 Type-1 and 5698 Type-2-galaxy pairs. The difficulties of detecting close neighbours due to spectroscopic fiber collisions causes pairs closer than $55''$ from each other not to be detected, unless residing in overlapping regions of the SDSS fiber plug plates. Therefore, photometric comparison samples are additionally constructed, with neighbours having photometric redshifts $|\Delta z| < 0.03$ and within a projected distance of 350 kpc. The final samples of galaxies with photometric redshift neighbours are 13519 Type-1 and 58743 Type-2 AGN-neighbour pairs. A cut in absolute magnitude for the neighbours (e.g. $M_r < -21.2$ as is used for the volume-limited samples below) removes stars and other faint objects among the contaminants.

The details of the sample selection, various selection effects and the treatment of biases predicted from clumpy dust torus models are all discussed in the Supplementary Information.

For the AGN-galaxy pair samples we study how the $u_e - r_e$ colour ("e" here stands for "internal extinction corrected") of the normal non-AGN neighbours behaves as a function of projected distance between the companion and the AGN. The colour of a galaxy can disclose information on the star formation, dust content, metallicity and age distribution of the stellar populations, i.e. important physical properties for understanding galaxy evolution. Besides removing the AGN among the neighbours, all the nuclear emission-line regions (LINERs) are removed.

Figure 1 shows that the average colour of the neighbours is redder around Type-1 AGN than around Type-2 AGN with $\sim 4.5\ \sigma$ significance. This finding points to systematic differences bewteen Type-1 and Type-2 AGN not captured in AGN unification. Instead, this could mean that neighbours to Type-1 AGN are experiencing less star formation, have more dust, a higher metallicity or older stellar populations than Type-2 AGN neighbours and that Type-1 and Type-2 AGN might have formed in different environments. The analysis is redone using two more $|\Delta z|$ cuts: $|\Delta z| < 0.001$ and $|\Delta z| < 0.006$ and the results stay consistent with the previous.

To improve the sample statistics and include the unseen neighbours due to fiber collisions, a similar analysis is done with the larger photometric neighbour samples, see Figure 2. The most striking feature is the very strong trend towards bluer neighbours very close to the Type-2 AGN, something that might indicate a strong increase in star formation or AGN activity in the neighbour.

It is conceivable that AGN unification is limited to some particular morphological type. In the Seyfert samples, Type-1 and Type-2 AGN hosts exhibit very similar trends as those in Figure 2. For the elliptical hosts, there are too few Type-2 neighbours to do a similar analysis, but for the closest bin we calculated the mean value for Type-2 AGN neighbours ($u - r \sim 2.02 \pm 0.04$) and for the Type-1 AGN neighbours ($u - r \sim 2.53 \pm 0.01$).

We also try to match Type-1 and Type-2 AGN better in stellar mass and AGN activity. The AGN unification predicts $L([\text{O\,{\sc iii}}])$ to be an isotropic indicator of AGN activity that measures the strength of the isotropically distributed narrow-line region (NLR) outside the torus. This means that Type-1 and Type-2 AGN with the same $L([\text{O\,{\sc iii}}])$ and redshift should have the same properties regarding their host galaxies (e.g. mass). We construct $L($[O III]$)$-matched control subsamples. Starting from the photometric neighbour samples, each Type-1 AGN is one-by-one matched with the Type-2 AGN having the closest redshift and $L($[O III]$)$. This includes also subsamples with only spiral host galaxies that are visually selected to be face-on in order to reduce dust extinction from the host galaxy plane. Therefore the same analysis is performed with the one-by-one matched control samples looking at colours of neighbour galaxies with the same absolute magnitude cut, $M_r < -21.2$. The results stay consistent (see Supp. Inf. section 1.7.2.).

We also calculate how the ratio between the number of Type-1/Type-2 neighbours around Type-2 AGN varies as a function of distance from the Type-2 AGN, see Figure 3. This ratio should not depend on distance, if Type-2 AGN as central objects do not prefer one type of AGN neighbours over the other. However, we find a clear increase (with 4.5 $\sigma$ significance) of the ratio at large projected separations. This is consistent with the observed deficit of Type-1 AGN in isolated galaxy pairs (16).

One could also wonder whether the presence of a neighbour might influence the morphology and AGN type of the host galaxy. It has been an open question whether there exists any correlation of host galaxy morphology with AGN type (17). Both questions can be addressed by comparing the morphologies of Type-1 AGN and Type-2 AGN in the parent samples with those having neighbours within two different projected distances: 100 kpc and 350 kpc. This allows us to examine how the presence of a nearby neighbour (indicator of future or past interaction/merger) alters the observed morphology of the AGN host galaxy, as mergers of gas-rich spiral galaxies of similar masses could become elliptical galaxies (18).

Figure 4 shows how the fraction of AGN in spiral host galaxies depends on AGN type and the presence of a neighbour. The clustering as seen by the number of galaxies in different $|\Delta z|$-cuts (see Supplementary Information, Table 2) suggests the two AGN types to reside in similarly dense environments. We note that the number of Type-1 AGN hosts with uncertain morphologies (neither spiral or elliptical) does not change significantly in the presence of a neighbour (see Supplementary Information, Table 4). For the case of Type-2 AGN, the fraction of host galaxies with uncertain morphologies decreases from 47.3% to 38.7% with 6 $\sigma$ significance – the morphologies of the host galaxies get much more well-defined in the presence of a neighbour. The behaviour of the host morphologies of Type-2 AGN as seen in Figure 4 contradicts the expectations from the morphology-density relation that predicts the fraction of spiral hosts to drop in denser environments.

The dramatic difference between Type-1 and Type-2 AGN host morphology behavior implies one of the following three scenarios: the companion has a much smaller mass than the Type-2 AGN favoring the formation of spiral arms in the AGN host galaxy (however, here the neighbours are fairly massive) *or* the Type-2 AGN do not merge with their neighbours (by some exotic mechanism) *or* the Type-2 AGN are fragile and are not preserved in their original state during/after merger. In the high-redshift universe when mergers were more common, the fragility would result in a deficiency of narrow-line AGN. This is indeed consistent with the observed lack of narrow-line quasars at high redshift.

So what could Type-2 AGN transform into during mergers? Elliptical Type-2 AGN might represent a form of transition objects between the two types and could explain why not all narrow-line AGN show broad lines in their polarized spectra (4; 19). We would in such a case expect to see that $\gg 50\%$ of the Type-2 ellipticals reveal a hidden BLR. This is testable by spectropolarimetric observations.

Evolution of intrinsic properties such as stellar ages with redshift could help to estimate the average life-time of the object. The Krongold-Koulouridis scenario (20) offers an evolutionary scenario where Type-2 AGN gradually transform into Type-1 and may agree with the decrease of the Type1/Type2 AGN ratio near Type-2 AGN and the increased number of blue, gas-rich neighbours near Type-2 AGN. In this scenario, the Type-2 AGN activity is initiated by the merger of two galaxies. At first, the star-formation induced by the merger will dominate the spectrum, but upon relaxation of the starburst and subsequent accretion of matter, a Type-2 nucleus is formed. When the merger completes and the AGN becomes so powerful that it blows away the dust torus, a Type-1 AGN is observable. The implied time scales for such a development would, however, require the AGN activity to be highly episodic in order for the super-massive black hole not to get heavier than what is observed.

We have demonstrated that the influence of active galaxies on close neighbours largely depends on the nature of the AGN – whether it is a broad-line or a narrow-line AGN. This is an unexpected result and shows the contrasting fates the two types of AGN face in mergers and interactions. With follow-up studies employing accurate host galaxy mass estimates, the missing links regarding the role of AGN in galaxy evolution may finally be disclosed once we accept the distinct voices the two classes of AGN have to play in the symphony of cosmic structures.

Both authors contributed equally to interpreting the results and writing the manuscript. The authors wish to thank Yoshiki Matsuoka and Bengt Gustafsson for invaluable feedback on the manuscript. B.V. wishes to thank Robert Antonucci for engaged discussions that helped to improve the manuscript, as well as Takis Konstantopoulos, Ernst van Groningen, Jesus Gonzales, Allen Joel Anderson, K. Szymanowski and Pianist. This work was funded and supported by the Center of Interdisciplinary Mathematics, the Crafoord foundation and the Swedish Royal Academy of Sciences. This research heavily relies on the Sloan Digital Sky Survey (SDSS). Funding for SDSS-II has been provided by the Alfred P. Sloan Foundation, the Participating Institutions, the National Science Foundation, the U.S. Department of Energy, the Japanese Monbukagakusho, and the Max Planck Society.

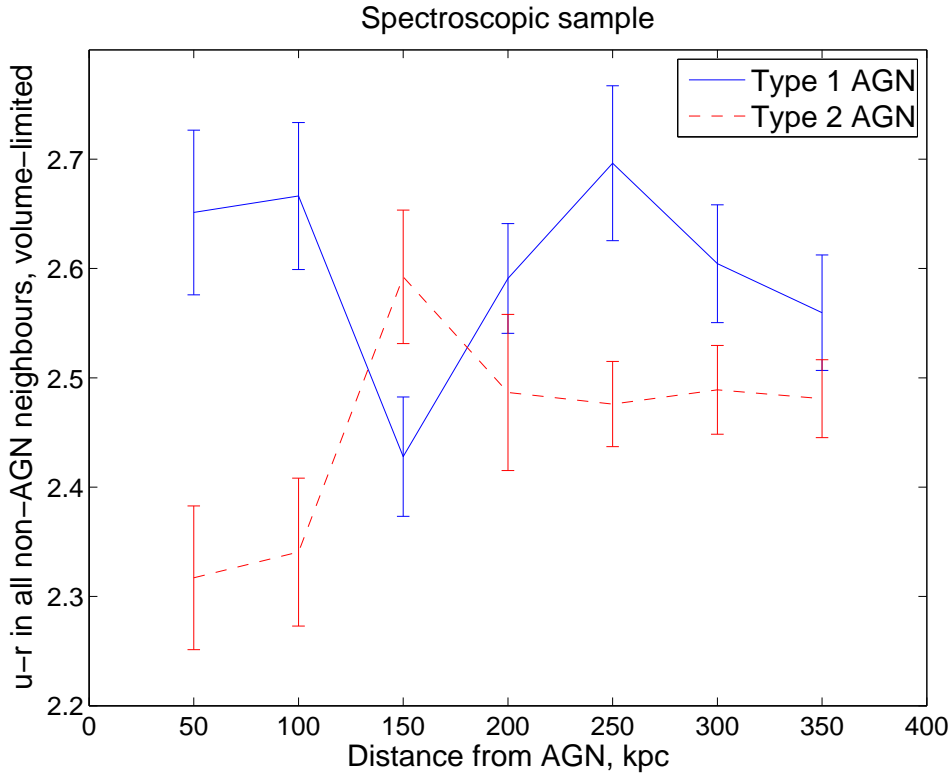

Fig. 1.— Mean $u_e - r_e$ colour of inactive neighbour galaxies. It is shown as a function of projected distance to the AGN. The data is grouped in 50 kpc bins. The spectroscopic redshift pairs have $|\Delta z| < 0.012$ and the neighbour colours are corrected for underlying stellar absorption and internal extinction ("e" in the colour indices stands for "internal extinction"). Volume-limited samples are used– both host galaxy and neighbour have $z < 0.14$ and $M_r < -21.2$. The number of pairs in the plot is 552 for the Type-1 AGN and 890 for the Type-2 AGN. Standard Gaussian errors for each bin are indicated by the $1\sigma$ error bars. The Type-1 AGN neighbours are redder on average than the Type-2 AGN neighbours with 4.5 $\sigma$ significance. The difference in colours of the neighbours disagree with the expectations from AGN Unification predicting the same colour of the neighbours. Changing the redshift difference cuts $|\Delta z| < 0.001$, 0.006 or 0.012 does not influence the average colours of the Type-1 and Type-2 AGN neighbours.

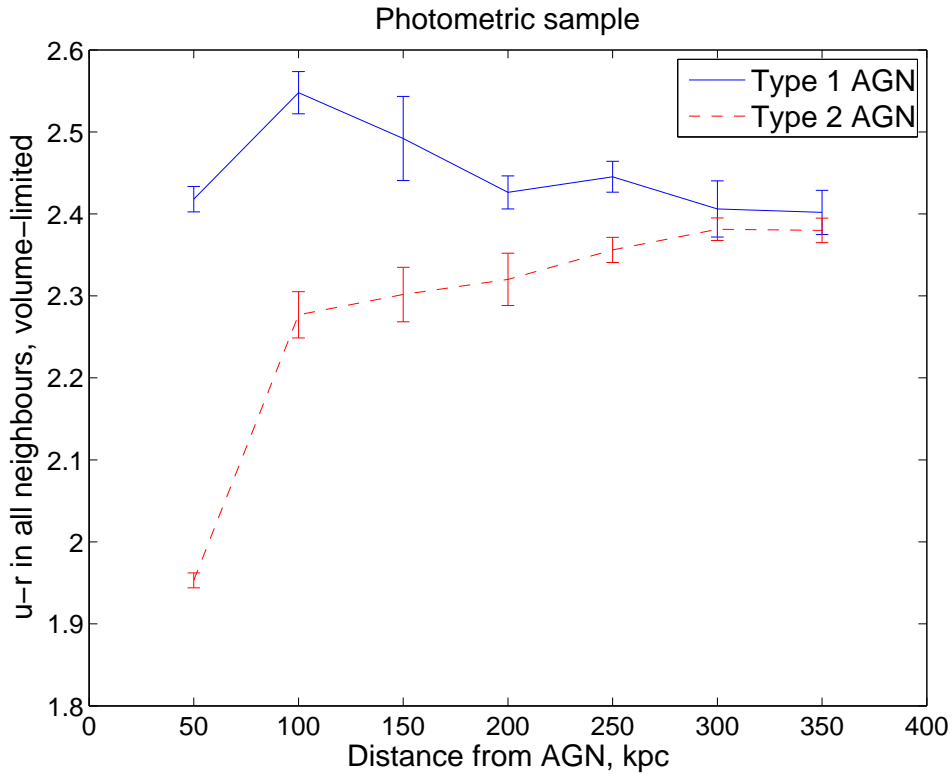

Fig. 2.— Mean $u - r$ colour of neighbour galaxies. It is shown as a function of projected distance to the AGN. The data is grouped in 50 kpc bins. Photometric redshift pairs that have $|\Delta z| < 0.03$ are used. Both host galaxy and neighbour have $z < 0.14$ and $M_r < -21.2$ (volume-limited). The number of pairs in the plot is 1836 for the Type-1 AGN and 3662 for the Type-2 AGN. Standard Gaussian errors for each bin are indicated by the $1\sigma$ error bars. Since no emission line information is available for these neighbours, we cannot correct them for internal extinction or remove the AGN among them. Therefore one can see that the colours are slightly bluer for these neighbours than those in the spectroscopic pair samples. The Type-1 AGN neighbours are redder on average than the Type-2 AGN neighbours with 20 $\sigma$ significance and show no correlation in the neighbour colour over projected distance. The Type-2 AGN neighbours become much bluer at the shortest projected separation with $\sim$ 24 $\sigma$ significance. The difference in colours of the neighbours disagree with the expectations from AGN Unification predicting the same colour of the neighbours.

Fig. 3.— Ratio in the number of Type1/Type2 AGN companions. The ratio of Type1/Type2 AGN with spectroscopic redshifts around Type-2 AGN as a function of distance from the Type-2 AGN. LINERs are excluded and we use neighbours with $|\Delta z| < 0.012$. Including LINERs among the companions does not influence the conclusions. In total, 92 Type-1 AGN and 527 Type-2 AGN companions. The errors are Poissonian and the significance level of the trend is 4.5 $\sigma$. The same trend is seen when using $|\Delta z| < 0.001$ (not shown here). In taking the ratio, many biases from sample selection will cancel each other. The trend disagrees with the expectations from the AGN Unification predicting a horizontal line. Changing the redshift difference cuts $|\Delta z| < 0.001, 0.006$ or $0.012$ does not influence the trend.

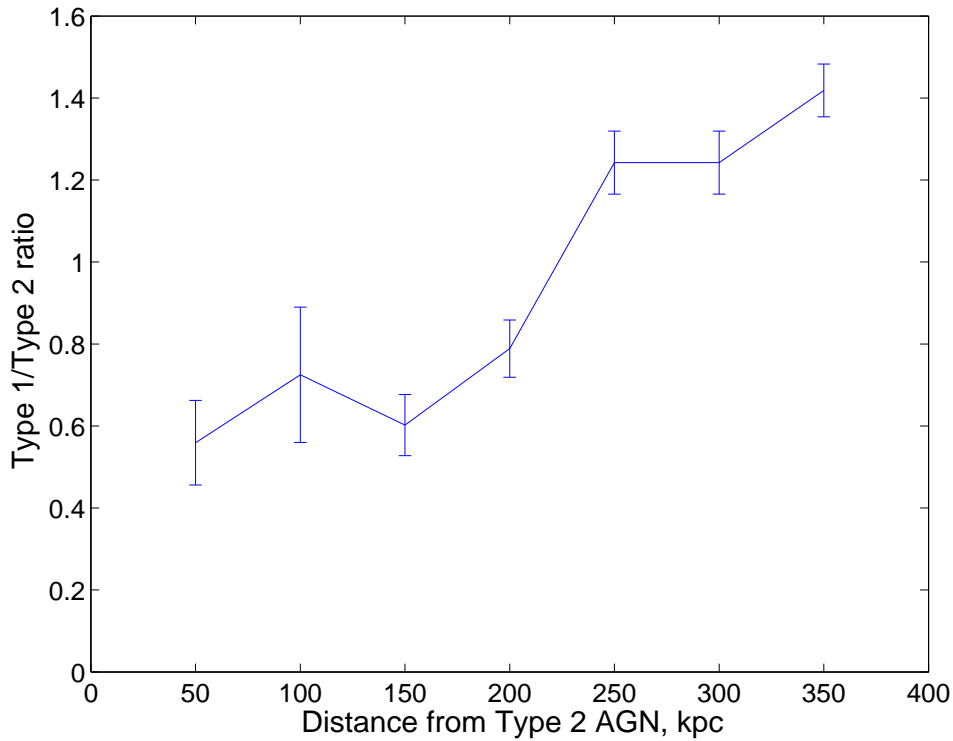

Fig. 4.— The fraction of host galaxies with spiral morphology in different environments. Both host galaxy and neighbour are in volume-limited samples with $z < 0.14$ and $M_r < -21.2$. We compare three different environments: AGN in the parent samples, AGN with neighbour within 350 kpc, and AGN with at least one neighbour within 100 kpc. The errors are Poissonian. Type-1 AGN residing in spiral hosts are 22% in the Parent Samples and decrease to 12% having a close neighbour within 100 kpc (7.5 $\sigma$ statistical significance when considering the difference of the two values). For Type-2 AGN, the fraction of AGN in spiral host galaxies on the contrary *increases* from 52% to 58% on a 2.5 $\sigma$ statistical significance level (or 52% to 59% (4 $\sigma$) for a neighbour within 350 kpc). This implies strong differences in how the morphology of Type-1 and Type-2 AGN host galaxy depends on the presence of a neighbour.

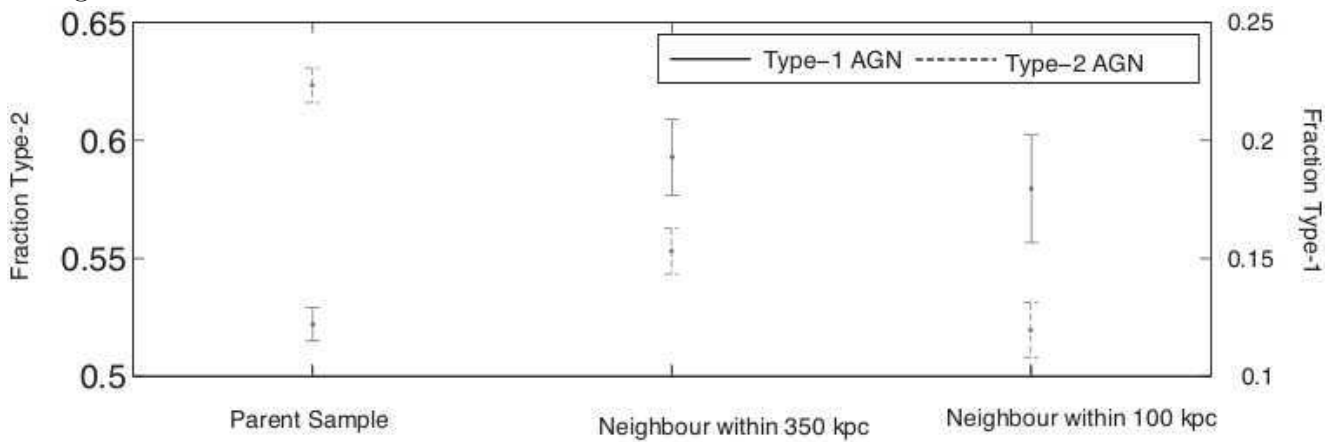

---

# 1. Sample selection

## 1.1. Spectroscopic sample selection

The SDSS (corrected for Galactic foreground extinction (1)) DR7 (2) has spectroscopic data of 929,555 galaxies and 121,363 quasars in five optical bands (*ugriz* filters). All objects were selected in the redshift range $0.03 < z < 0.2$. We used objects with specClass= 2 or 3 ('Galaxies' or 'Quasars'), $EW(H\alpha) > 2$ Å and reliable redshifts ($z$conf >0.95). We note that the AGN we use come from both specClasses, as can the neighbours. We rejected those with poor quality indicated by catalog flags: brightness (flags&0x2=0), saturation (flags&0x40000=0) and blended images (flags&0x8=0) (3, see Table 9). The reason why we reject objects with $EW(H\alpha) < 2$ Å is that our selection methods for Type-2 AGN is purely based on emission-line criteria and we therefore must restrict our Type-1 AGN sample to emission-line objects only for consistency. As a consequence, we remove a small number of absorption-line objects such as BL Lac objects. (However, the majority of all AGN are emission-line objects.)

As we in an earlier publication (4) lost 5% of the quasars due to flags, we consider the bias from flagging being a negligible effect that will not significantly influence the number of AGN.

We retrieve redshifts, spectral line information and Galactic foreground extinction-corrected apparent magnitudes ('dered'), resulting in 253,352 objects from which we derive the parent AGN samples. For the objects marked as 'Galaxies' we retrieve K-corrections from the PhotoZ table. For the objects marked as 'Quasars', we calculate the K-correction using a universal power-law spectral energy distribution (SED) with optical flux given by $f_\nu = \nu^\alpha$ where the mean spectral optical index $\alpha=-0.5$ can be related to the K-correction using:

$$K(z) = -2.5\alpha \, log(1 + z) - 2.5 \log(1 + z) \quad (1)$$

The Galactic extinction- and K-corrected rest frame absolute magnitudes of all objects can then be calculated according to:

$$M_{abs} = m_{obs} + 5 - 5\log(D_L) - K(z) \quad (2)$$

Table 1 summarizes all the samples in our study and their definitions.

### 1.1.1. LINERs

Low-ionization nuclear emission-line regions (LINERs) occur in a relatively large fraction of nearby galaxies of different morphologies and luminosities. These regions have an emission-line spectrum with low-ionization states. Emission lines from higher ionization stages are weak or absent. LINERs are the subject of scientific debate, since it is unclear whether these nuclear emission regions arise from AGN activity or from star-forming regions. Neither is the mechanism behind the low ionization known, and both shock waves and UV light are argued to be the main reason. LINERs are commonly referred to as AGN in scientific literature. But there are observations that indicate that the AGN Unification might break down by disappearance of the torus or broad-line region in low-luminosity AGN (5; 6). This suggests that it might be the best to actually separate the LINERs from the main AGN samples in order to study the AGN Unification for objects where it is more likely to be true.

However, we separate them in our study with the help of the definition, the Kewley criterion (7):

$$\log([O\,III]/H\beta) > 0.61/(\log([N\,II]/H\alpha)) - 0.47) + 1.19 \quad (3)$$

Using this criterion, we obtain 34249 LINERs in our parent sample. Here we will only use them to compare the redshift distributions of the three types of AGN. The selection criterion is mainly be used in this work to remove the LINERs from the Type-1 and Type-2 AGN samples. We define this sample as the "LINER Parent Sample".

As neither the influence of AGN nor LINERs on the star-formation of galaxies is well-established, we also use this criterion to remove LINERs from the spectroscopic neighbour samples. This is necessary if one wishes to investigate the potential influence from AGN on their neighbour properties (e.g. star-formation rate or colour) and clearly separate it from effects caused by an increased LINER fraction.

### 1.1.2. Type-1 AGN

Broad Balmer emission lines are the key signature of an accretion disk. The Type-1 AGN were therefore selected on solely one criteria, the width of the $H\alpha$ line in the SDSS single-Gaussian fitted spectra, measured in the emitter's rest frame.

Galaxies should fulfil spectral line width $\sigma(H\alpha) > 10$ Å (corresponding to FWHM > 1000 km s$^{-1}$) and emission in $H\alpha$ in order to be categorized as Type-1 AGN. LINERs were excluded by using the Kewley criterion. The first selection results in 11334 Type-1 AGN. We define this sample as the "Type-1 Parent Sample".

The SDSS single-Gaussian fitted spectra can bias the separation between Type-1 and Type-2 AGN, since it might be difficult to resolve the contribution to the $H\alpha$ emission-line peak from the [N ] contribution from a narrow-line AGN. Type-2 AGN can also get "mixed" into the sample in case the broad-line component is faint, if we have scattered light from a broad-line region in an obscured quasar or if the forbidden lines in a Type-2 AGN would have a non-Gaussian line profile (8). We address the first of these three problems later in this work by comparing neighbours depending on the S/N ratio of the $H\alpha$ line and equivalent widths of the AGN hosts.

When experimenting with a different limit, $\sigma(H\alpha) > 15$ Å to select the Type-1 AGN, the same analysis results were obtained.

### 1.1.3. Type-2 AGN

To select narrow-line AGN we use the Baldwin-Phillips & Terlevich line-ratio diagrams (9) combined with the Kauffmann (10) criterion:

$$\log([\text{O{\sc iii}}]/\text{H}\beta) > 0.61/(\log([\text{N{\sc ii}}]/\text{H}\alpha)) - 0.05) + 1.3 \quad (4)$$

To avoid contamination of Type-1 AGN in the Type-2 AGN Parent Sample, we only include Type-2 AGN having $\sigma(\text{H}\alpha) < 10$ Å. These include many composite objects also hosting starburst activity. LINERs were excluded by the use of the Kewley criterion, see above. The first selection results in 53416 Type-2 AGN, defined as the "Type-2 Parent Sample". We will also for the Type-2 AGN sample check the impact of the S/N ratio of the emission line on the conclusions.

### 1.2. Volume-limited test sample of parent objects

We chose objects with $M_r < -21.2$ and $z < 0.14$ in order to get the largest possible number statistics (with respect to the number of Type 1 AGN) for a volume-limited subselection of our Type-1 AGN, Type-2 AGN and LINER Parent Samples. The redshift distributions of the three volume-limited subsample can be seen in Figure 1. It is in fair agreement with the redshift distribution of other samples (11; 12) and the narrow-line AGN sample selected with the Kauffmann criterion.

Our parent samples contain approximately 3.5 times as many Type-2 AGN as Type-1 AGN in this volume-limited subsample, see Table 1. This is not in contradiction with other studies where the ratio for Type-1 to Type-2 AGN varies between 1:2 to 1:5 (13). We also include the absolute magnitude distribution for the same samples, see figure 2.

AGN unification also predicts $L([\text{O }])$ to be isotropically distributed in Type-1 and Type-2 AGN. The big overlap in $L([\text{O }])$ that most studies find have been a strong support for the model, and some works have found an almost precise agreement in the $L([\text{O }])$ (14).

We check that our volume-limited parent samples are more or less similarly distributed in $L([\text{O }])$, and conclude that the two parent samples of Type-1 and Type-2 AGN are fairly similar albeit not identical, see Fig. 3. The mean values in $L([\text{O }])$ is $39.7386 \pm 0.0157$ for Type-1 vs $39.5261 \pm 0.0043$ for Type-2.

We could ask ourselves whether this difference would not have been noticed in smaller samples too, as in e.g. Keel et al. (1994). In their study, they used 80 galaxies with Seyfert 1 spectra and 141 galaxies with Seyfert 2 spectra. For a fair comparison with respect to the sample sizes, we randomly (with MATLAB's data-sample algorithm) select a subset of 80 galaxies from the Type-1 parent sample and a subset of 141 galaxies from the Type-2 parent sample from those displayed in Fig 3. The resulting mean values in $L(\text{O })$ is $39.5748 \pm 0.1203$ for Type-1 vs $39.4722 \pm 0.0521$ for Type-2. The difference between these two samples is less than one sigma. This might explain Keel at al.'s conclusion that their Seyfert 1 and Seyfert 2 samples were isotropic in $L([\text{O }])$, and suggests that for the same sample size, our Type-1 and Type-2 AGN are at least as similar to each other in $L([\text{O }])$ as the samples in Keel et al. (1994).

### 1.2.1. Neighbours to AGN

Since surveys have a limited redshift accuracy we must work with projected distances instead of real physical distances when we select neighbours. This means that a lot of background and foreground objects are expected to be sampled even when using the smallest redshift difference cut $|\Delta z|$ as this cut will correspond to much larger distances at the Mpc scales. When we investigate the behavior of neighbours over our selected projected distance of 0–350 kpc, we therefore assume that these background objects are isotropically spread over the projected distance range and therefore cannot influence the results and conclusions or create any false trend. If the conclusion would depend on our choice of $|\Delta z|$, we would know the results would not be robust. Therefore, we investigate all our results with different choices in $|\Delta z|=0.001, 0.006, 0.012$ and $0.03$ and confirm they stay the same independent of the choice. This will be of crucial importance to know once we will use neighbours with photometric redshifts were the accuracy is pretty low.

To select neighbours to our AGN in the parent samples, we query for all nearby galaxies from the SpecObjAll catalogue (DR7) to our AGN within 11 arcminutes with the function "dbo.fDistanceArcMinEq" and within $|\Delta z| < 0.012$ from the main AGN. This angular separation was selected in order to avoid a biased sample within the projected distance range of interest, which in this work is between 0 to 350 kpc.

We are interested in the 0–350 kpc range as it covers both close interacting pairs as well as the nearby large-scale environment, where interesting phenomena – such as a sudden gap in the surface density of blue neighbours to quasars at a distance around 150 kpc (4) – may occur. The redshift difference cut $|\Delta z| < 0.012$ is chosen as it provides us with the interesting opportunity to probe the large-scale environment in the redshift dimension, as we can investigate the clustering of galaxies by comparing the number of pairs within $|\Delta z| < 0.001$ (at $z=0.2$ corresponding to 4 Mpc), $|\Delta z| < 0.006$ (26 Mpc) and $|\Delta z| < 0.012$ (53 Mpc). Also, it has the benefit of being comparable to the photometric redshift neighbour sample as this value is roughly half of the photometric redshift error, $\delta z < 0.025$.

After removing repetitions, we retrieved 1658 Type-1 AGN-galaxy pairs (defined as the "Type 1 Spectroscopic Pair Sample"), 5698 Type-2 AGN-galaxy pairs ("Type 2 Spectroscopic Pair Sample") and 4214 LINER-galaxy pairs ("LINER Spectroscopic Pair Sample"). For these we obtain the dereddened apparent magnitudes, the spectroscopic redshifts, the isophotal axis lengths, the spectral line information, K-corrections from the PhotoZ catalogue and calculate the projected distances between the two object in each pair. We also calculate the rest-frame absolute magnitudes for each object as described in Section 2.1.

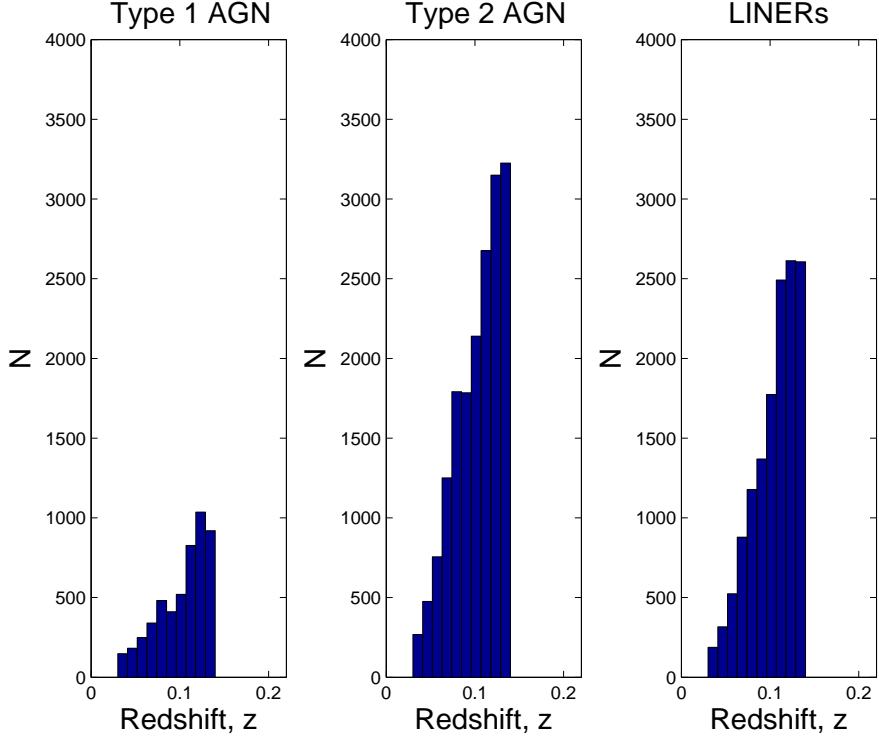

**Fig. 1.** Distribution of redshift in the volume-limited parent AGN samples. Median redshift is z ∼ 0.13 for Type-1, z ∼ 0.09 for Type-2 and z ∼ 0.11 for LINERs.

We investigate the $|\Delta z|$ distributions of our pairs. If the objects within the pairs selected within $|\Delta z| < 0.012$ would all be physically unassociated, the distribution of the number of pairs over different, equidistant $|\Delta z|$ bins ($|\Delta z|$= [0, 0.001], [0.001, 0.002],...) would be more or less uniform. We investigate the clustering and notice that the distribution of pairs sharply peaks at $|\Delta z| < 0.001$ for the spectroscopic neighbour samples and then rapidly falls off, see Table 1. This means, that most of the objects within the pairs are physically associated, as interacting pairs that are selected on tidal distortion would seldom have $|\Delta z| > 0.003$.

#### 1.2.2. Photometric redshift neighbours

Due to spectroscopic fiber collisions approximately ∼ 2/3 of all pairs with angular separation less than 55″ can not be detected (15), unless residing on overlapping spectroscopic plug plates. This might bias our sample towards more wide pairs. The finite fiber size also makes it difficult to detect pairs in the late stages of mergers.

To get a more valid estimate on the clustering of different colour types of galaxies around Type-1 and Type-2, we query the PhotoZ catalogue to find all neighbour galaxiesq with photometric redshifts. We do this only for the AGN already existing in our spectroscopic AGN-galaxy pair catalogues and select all neighbours within the angular distance of 11 arcmin and a $|\Delta z| < 0.03$ from the AGN.

While the photometric redshift samples do not suffer from the fiber incompleteness issue, the resolution and surface brightness limits of the SDSS photometry might play a role in the detection of neighbours. Faint neighbours may not be detected, and if any AGN has an increased number of low surface brightness (LSB) galaxies at short projected separations, these might not be observed and hence bias our sample towards an underestimated clustering of LSB galaxies around the AGN. For the photometric pairs, we retrieve the morphologies for all AGN hosts from the Galaxy Zoo project, see below. Our final samples (including morphologies) are 13519 Type-1 AGN-galaxy pairs (defined as "Type 1 Photometric Pair Sample") and 58743 Type-2 AGN-galaxy pairs (Type 2 "Photometric Pair Sample"). The ratios of pairs between the photometric pair samples and spectoscopic pair samples indicate that the Type-2 AGN have more nearby neighbours than can be detected by SDSS spectroscopy (ratio spectroscopic/photometric ∼ 0.38 for the Type-1 morphology neighbour samples, the same ratio for Type-2 is ∼ 0.33)

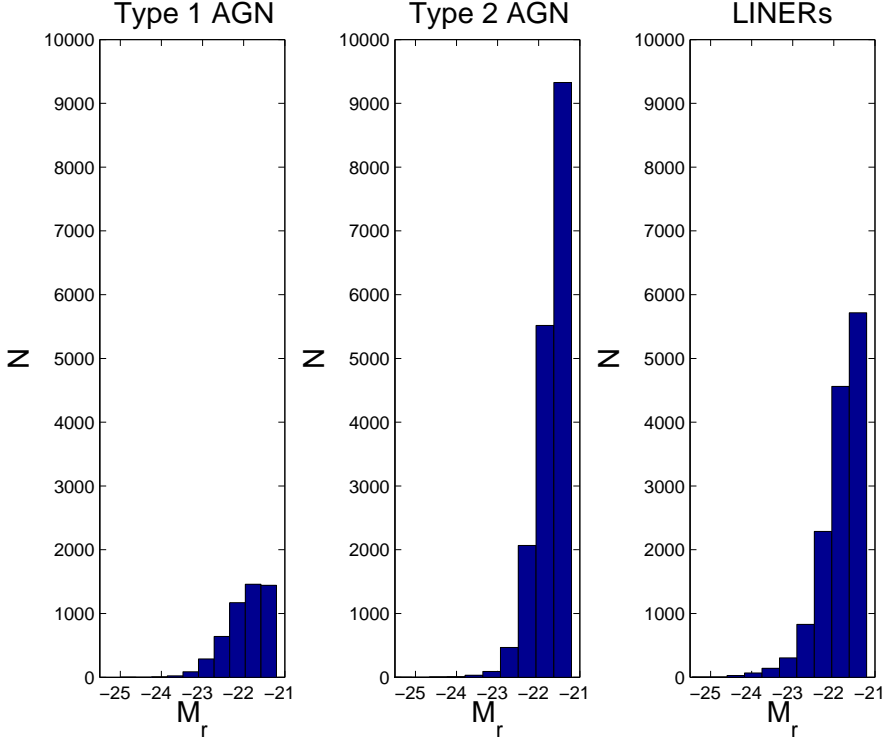

**Fig. 2.** Distribution of $M_r$ in the volume-limited parent AGN samples. Median absolute magnitude is $M_r \sim -21.9$ for Type-1, $M_r \sim -21.6$ for Type-2 and $M_r \sim -21.7$ for LINERs.

Together with the information from $|\Delta z| < 0.001$, 0.006, 0.012 and 0.03 we cover the transition region between the spectroscopic and photometric neighbour samples and can thus discover potential effects in the neighbours caused by background galaxy contamination.

The computed errors in photometric redshifts are large at low redshift $z < 0.1$. We investigate how the two main figures Fig. 2 & 4 (Letter) are influenced if we only use neighbours with computed photometric redshift errors $|\delta z| < 0.03$. The results stay robust, despite that we lose approximately 6−7% of the pairs.

#### 1.2.3. Morphology samples

For the pairs, we chose to query for the morphology of the parent, AGN hosts in the pair from the Galaxy Zoo project (16; 17). Not all parent AGN in our pair samples had morphology information, but those that had are in the "Morphology Parent Samples". This is why the morphology samples are very similar to, but slightly smaller, than the spectroscopic redshift samples. They are defined as "Morphology Pair Samples". We also obtain the morphology classifications for our parent AGN samples, so that we can see if the presence of a close neighbour significantly changes the morphology of the AGN hosts. The samples are summarized in Table 4.

#### 1.3. Line flux and extinction corrections

The spectral line measurements and colors of the non-AGN neighbours are corrected for underlying stellar absorption and internal dust extinction in the spectroscopic neighbour samples as in Villarroel 2012 (4). For the neighbouring galaxies to the AGN in our spectroscopic redshift pairs, we correct the Balmer emission line fluxes and equivalent widths for underlying stellar absorption by assuming average absorption line strengths corresponding to 2.5 Å in $EW$ for H$\alpha$ and 4 Å for H$\beta$ (18). Further, we perform internal extinction corrections for neighbour galaxies with $EW(H\beta) > 5$ Å according to a standard interstellar extinction curve (19; 20). Here the stellar absorption-corrected H$\alpha$ and H$\beta$ lines for each galaxy are used to calculate the extinction coefficient which together with the interstellar extinction curve can be used to estimate the amount of extinction for each emission line.

The colours of the neighbours were corrected for inclination-dependent dust extinction using analytical expressions (21), meaning that we corrected the colours for all galaxies having $0 <$ $(u - r) < 4$, H$\alpha$ line width ranging from 0 to 200 Å and absolute magnitude $-21.95 < M_r < -19.95$ and finally a concentration index C in the range $1.74 < C < 3.06$. Neighbour galaxies outside these parameter ranges are left untreated, which for instance means that more luminous objects have no inclination-dependent extinction correction applied to their colours. Unless all neighbour galaxies around only one certain type of AGN would be observed with the same inclination angle (which is highly unlikely due to the assumed isotropy of the Universe) *or* the selection of the two AGN types has strong biases, we do not expect extinction correction to significantly change the results. We control this by not including any extinction corrections on colours and emission lines, which turns out not to change the results of this work at all and the applied extinction corrections are therefore merely used for improved accuracy.

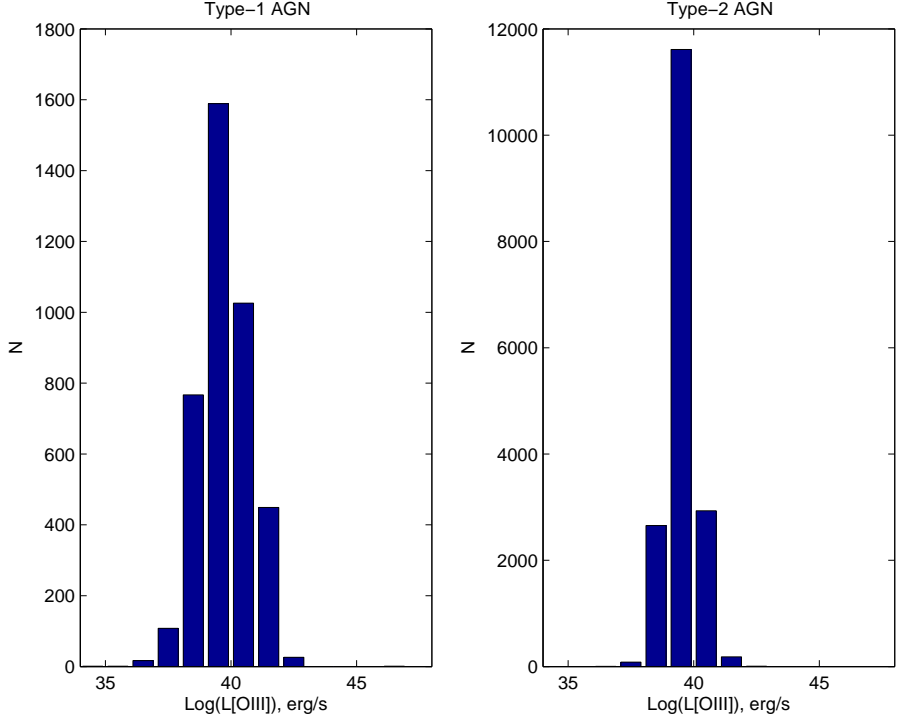

**Fig. 3.** Distribution of $L([O\,\textsc{iii}])$ in the volume-limited parent AGN samples.

**Table 1.** A summary of all the samples in our study. The number of objects in the volume-limited cut of the sample is given ($z < 0.14$ and $M_r < -21.2$). The total number of objects in each sample is given in parenthesis.

| Samples | | | |
|---|---|---|---|
| Sample type | Type 1 | Type 2 | LINER |
| Parent | 5114 (11334) | 17518 (53416) | 13939 (34249) |
| Spectroscopic Pair | 703 (1658) | 1194 (5698) | 1282 (4214) |
| Morphology Parent | 5065 (10154) | 16294 (48970) | - |
| Morphology Pair | 702 (1604) | 1191 (5547) | - |
| Photometric Pair | 1836 (13519) | 3663 (58743) | - |

### 1.4. Volume-limited test sample of pairs

We estimate the relative fraction of pairs by comparing the spectroscopic pair samples to the spectroscopic parent samples.

From the second column of Table 2, we can not see any significant difference in $|\Delta z|$ between the different pair samples. On the other hand, we do see that the Type-1 AGN significantly more often have a close neighbour than do Type-2 AGN or LINERs. Our results are in direct contradiction with the companion-count results from works (22; 23) that report a larger number of neighbours around Type-2 AGN than around Type-

**Table 2.** Number of satellites around central AGN in a volume-limited pair samples. The number of satellites in the subsample $|\Delta z|=0.001$ (column 2) is compared to the Parent Sample $|\Delta z|=0.012$ (column 4). The fraction of satellites in the subsample $|\Delta z|=0.001$ (column 3). The ratio (column 5) between the number of satellites in subsample $|\Delta z|=0.001$ and total number of objects in a volume-limited pair sample (column 6).

| Clustering of neighbours | | | | | |
|---|---|---|---|---|---|
| Central galaxy type | $|\Delta z|=0.001$ | fraction$_{\Delta z}$ | $|\Delta z|=0.012$ | fraction$_{pairs}$ | N Parent Sample |
| Type-1 | 385 | 0.55 | 703 | 0.08 | 5114 |
| Type-2 | 718 | 0.60 | 1194 | 0.04 | 17518 |
| LINER | 794 | 0.62 | 1282 | 0.06 | 13939 |

1 AGN. However these particular works used galaxies at lower redshift than ours and could possibly have detected many more low-surface brightness galaxies around the Type-2 AGN. This would also agree well with the differences in the ratio of companions between the volume-limited and non-volume-limited samples.

For the AGN themselves we also compare the absolute magnitude $M_r$, rest-frame colour $u_e - r_e$ and redshift distributions to AGN in the parent samples. No significant changes in luminosity, colour and redshift, depending on the presence or absence of a companion, are found.

### 1.5. Star formation rate

We use the measured H$\alpha$ emission line strength from the spectroscopic catalog of the SDSS together with the Bergvall-Rönnback calibration (1995) to estimate the star-formation rate in the neighbours of AGN.

$$L(H\alpha) = SFR * 1.51 * 10^{34} \qquad (5)$$

and

$$L(H\alpha) = 4\pi D_L^2 \sqrt{2\pi}\sigma h 10^{-20} \qquad (6)$$

where SFR is the star formation rate in solar masses per year, $\sigma$ and $h$ are width and height of the H$\alpha$ emission line, $D_L$ is the luminosity distance in Mpc and the emission line luminosity is expressed in Watts.

We first try to see if we can find any correlation or differences of SFR with distance between AGN and neighbour within the volume-limited cuts. The small number of Type-1-galaxy pairs (153) and Type-2-galaxy pairs (365) within the volume-limited cut does not permit us to detect any statistically significant differences in the neighbour samples. We therefore search for an indication by estimating the number of neighbours (see Table 3) within this volume-limited cut that have measurable star formation. In this study, we define the star formation rate as "measurable" if the flux in H$\alpha$ > 0 and the flux is twice the flux error. We find that 31 % of the Type-2 AGN neighbours have measurable star formation rate, while only 22 % of the Type-1 AGN neighbours do, with a statistical significance level of 3.3$\sigma$. This could suggests that the Type-2 AGN either form in regions with abundant gas supply or that they could transfer gas to the neighbour galaxies. Any of these explanations would support the observed higher column densities of molecular hydrogen in the Type-2 AGN (25). Improved statistics on neighbour galaxies with measurable star formation rate would be needed to confirm (or refute) this hypothesis.

One could also argue that Type-2 AGN are dustier objects on average, and therefore suppress the star formation rate of the neighbours less. We in addition notice that the Type-1 AGN neighbours have a higher dust content (measured in H$\alpha$/H$\beta$-ratio) than the Type-2 neighbours, but since the H$\alpha$ line is corrected for underlying stellar absorption, Galactic and internal extinction, this is most likely not the reason for the observed differences in star formation rates.

### 1.6. A Hypothetical Luminosity Test of AGN unification

As the intrinsic properties of AGN are unknown due to the lack of knowledge on the geometry of the dust torus, the high discrepancies between the neighbour populations of Type-1 and Type-2 AGN could be the result of a biased selection on bolometric luminosities and masses of the objects. There is an increasing evidence for that the torus also might be clumpy (26; 27).

It is very difficult to estimate how much the torus on average obscures the Type-2 AGN in the case of Realistic AGN unification (28; 29) due to the little observational evidence. The only thing we know is that if an AGN would be obscured by a dust torus, it would appear less luminous and could be mistaken for being less massive. This means that Type-2 AGN of lower luminosities could be considerably more massive than Type-1 AGN of the same luminosities.

One obvious way to approach the problem of a possible mass bias would be to extract the stellar and black hole masses of the central AGN in the two samples. We, however, chose to approach the problem by matching the neighbour samples in colour, while estimating the luminosity displacement in the Type-2 AGN relative to the Type-1 AGN, assuming a Gaussian distribution of the covering factor.

The risk of doing this for two random samples of Type-1 and Type-2 AGN is that in case the accretion disk of the Type-1 AGN outshines the host galaxy, one could end up comparing the luminosity from the accretion disk in Type-1 AGN to the luminosity of the host galaxy in Type-2 AGN. The advantage of using low-redshift AGN for this study becomes clear in the con-text; the assumingly equal contribution (under the assumption of AGN unification) of the host galaxy luminosity in rather faint low-redshift AGN makes it possible to compare the luminosity displacement on a statistical level, without any concerns for that the accretion disk of the Type-1 AGN could outshine the the host galaxy as in high-redshift AGN. As a result we are able, as a check of consistency, to investigate if there exists any luminosity displacement where the the two classes of AGN have similar neighbours (and thus assumingly similar masses). If they further display the same distributions of morphologies among host galaxies, then the two necessary conditions for AGN unification are fulfilled.

**Table 3.** The number of neighbours with measurable SFR, defined as blue and defined as AGN. |Δz|=0.012 is used.

| Central galaxy type | measurable SFR | fraction$_{SFR}$ | blue | fraction$_{blue}$ | AGN | fraction$_{AGN}$ | Total number of pairs |
|---|---|---|---|---|---|---|---|
| Type-1 | 412 | 0.25 | 527 | 0.32 | 243 | 0.15 | 1658 |
| - volume-limited | 153 | 0.22 | 119 | 0.17 | 110 | 0.16 | 703 |
| Type-2 | 2228 | 0.39 | 2323 | 0.41 | 1001 | 0.18 | 5698 |
| - volume-limited | 365 | 0.31 | 265 | 0.22 | 238 | 0.20 | 1194 |

**Table 4.** The Galaxy Zoo morphology of host galaxies in volume-limited samples. Both host galaxy and neighbour have $z < 0.14$ and $M_r < -21.2$. Two upper rows: The morphology of host galaxy in spectroscopic parent samples. Middle rows: The morphology of host galaxies (spectroscopic) with at least one neighbour (photometric) within 100 kpc. Bottom rows: The morphology of host galaxies (spectroscopic) with at least one neighbour (photometric) within 350 kpc. Not all AGN hosts had morphologies from the Galaxy Zoo. Those with "uncertain" morphologies can be calculated by substracting the number of spiral and elliptical from the total number of hosts. Errors in mean redshift are $\delta_z \sim 0.01$ and indicate the standard error. The mean redshift is the same for all three types of host morphologies. To calculate the statistical significance of the change in the fraction of morphologies between the samples, we use Poisson errors.

| Central galaxy type | spiral hosts | elliptical hosts | uncertain hosts | mean redshift | total number of hosts |
|---|---|---|---|---|---|
| Type-1$_{parent}$ | 1128 (22%) | 751 (15%) | 3186 (63%) | 0.123 | 5065 |
| Type-2$_{parent}$ | 8506 (52%) | 81 (<1%) | 7707 (47%) | 0.098 | 16294 |
| Type-1$_{100kpc}$ | 116 (12%) | 296 (30%) | 563 (58%) | 0.088 | 975 |
| Type-2$_{100kpc}$ | 1008 (58%) | 28 (2%) | 703 (40%) | 0.071 | 1739 |
| Type-1$_{350kpc}$ | 280 (15%) | 449 (25%) | 1107 (60%) | 0.088 | 1836 |
| Type-2$_{350kpc}$ | 2172 (59%) | 70 (2%) | 1421 (39%) | 0.071 | 3663 |

We herein propose a test to sort out whether a luminosity displacement is enough to explain the discrepancies.

### 1.6.1. Step one: determining the variables

While the general clustering of galaxies around the central AGN is rather similar for the two populations, we have seen that the pairs mainly differ in the following properties:

1. Clustering of galaxies with measurable star formation around the central AGN.
2. Average colour of the neighbour population.
3. Correlation of colour with distance between neighbour and AGN.
4. Morphology of the AGN host galaxy.

We here define the luminosity displacement in the AGN, $E_{dis}$

$$E_{dis} = M_{r,tot} - M_r \qquad (7)$$

We define an average "offset" ($O_{colour}$) in the colour $u - r$ between the neighbour populations of the Type-1 and Type-2 populations for projected distances d > 50 kpc. The reason for this choice is that a potential Holmberg effect (30), the possibility that more star-forming galaxy neighbours tend to be aligned perpendicular to the disk of the galaxy, could influence the star formation rate at short separations (d < 50 kpc), but would be unnoticed at larger radii. However, in this particular case, we would also see the same distribution of AGN host morphologies among Type-1 and the luminosity-displaced Type-2 AGN.

### 1.6.2. Step two: starting sample

The starting point consists of two volume-limited subsamples of both Type-1 and Type-2 AGN, where $M_r < -21.2$ and $z < 0.14$ is used for both central AGN and neighbour galaxies. While $M_r$ is held constant for the neighbour galaxies and the Type-1 AGN, we vary this parameter for the central Type-2 AGN until we reach a state where the offset will be close to $O_{colour} \sim 0$ and both neighbour populations are very similar to each other. We define this point as the first $M_r$ value where the 1 $\sigma$ error bars of the two sample overlap.

We perform the test for the two extreme cases to probe possible hidden obscured Type-1 AGN among the Type-2 AGN, the first as narrow as the volume limited Type-1 AGN sample in

**Fig. 4.** Results of the hypothetical luminosity test. The x-axis shows the lower cut in absolute magnitude both for the one-cut and the fixed-width luminosity test. Y-axis shows the K- and Galactic extinction corrected colours among the neighbours. A "candidate" luminosity displacement is found where the error bars overlap.

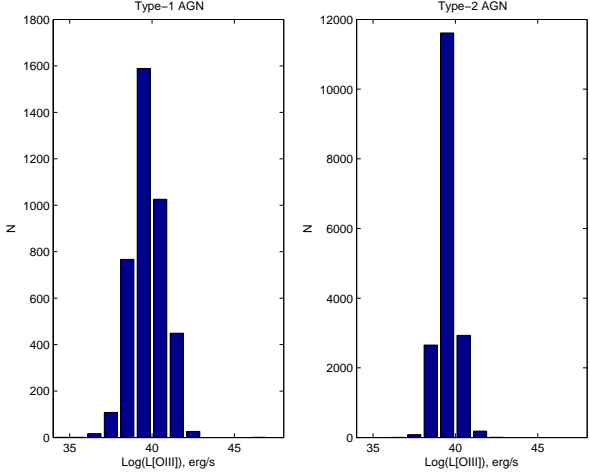

magnitude ("Single magnitude cut"), the second infinitely broad ("Double magnitude cuts"):

1. Varying the lower luminosity cut $M_r < -21.2$, where we iterate it in the direction $M_r > -21.2$ as far as we have data.
2. Varying the lower luminosity cut $M_r < -21.2$, where we iterate it in the direction $M_r > -21.2$ as far as we have data, plus an upper cut $M_{r,\text{upper}}$ where the range is equivalent to the fix width of the volume-limited distribution of Type-1 AGN in the pair sample. Since the most luminous Type-1 AGN in our volume-limited pair sample have an absolute magnitude $M_{rmax,\text{Type-1AGN}} = -23.8$, this gives us a comoving, fixed width of w=2.6 mag.

A support for AGN unification would occur during the iterations if $O_{colour} \sim 0$, and the new, deduced correlation of and colour (Figure 2, Letter) with projected distance are fairly similar between the two populations and the Type-2 AGN pair subsample with $O_{colour} \sim 0$. A second condition is that the new suggested Type-2 luminosity population must have roughly equal fractions of spiral and elliptical AGN among the hosts.

The average redshift of two populations with similar offsets can also reveal a minor time-evolution that otherwise can go unnoticed. The same test will also be used in future publications to connect LINERs to the evolution of active galaxies.

### 1.6.3. Test results

Figure 4 shows the results indicating that a subsample of Type-2 AGN within the luminosity range of $-17.2 < M_{r,\text{Type-2AGN}} < -19.8$ (corresponding to a luminosity displacement $E_{dis} \sim 4$ mag) could correspond in mass to the volume-limited Type-1 in the pair samples. Using only a lower cut did not influence the selection. The next step was to redo the colour-distance plot (Figure 2, Letter) with these luminosity constraints imposed on the Type-2 host luminosity selection and investigate the relative fractions of the host morphologies. The plot revealed that if selecting Type-2 AGN with the magnitude range $-17.2 < M_{r,\text{Type-2AGN}} < -19.8$, the Type-2 AGN neighbours turn out to be very similar to the Type-1 AGN neighbours in their colour-distance behavior. However, the distribution of morphologies were similar to those in the volume-limited parent sample, see Table 1 (Letter). In this selection among the Type-2 AGN with neighbours, 25% could be detected in spiral hosts and only 2% in elliptical hosts, while for the parent sample the corresponding number was 52% in spiral hosts and 1.5% in elliptical. Even if the different covering factors of the torus are insufficient to explain the observed difference in morphology populations, the similarity in neighbours of Type-1 AGN with more faint Type-2 AGN could indicate an evolutionary link between the two types of AGN, where the change of morphology during interaction seems to be one of the key ingredients.

We redid this test with the sample of spiral AGN, but here no magnitude range or cut could set the colour offset $O_{colour}$) to zero.

Additionally, we investigated how the S/N ratio in any of the diagnostic lines to select the AGN could influence our results by following the same principle as above but with S/N ratio on the x-axis and the same magnitude limits for both samples $M_r < -21.2$ (both for AGN and its neighbour). We did this for all emission lines separately for both Type-2 AGN and the Type-1 AGN's H$\alpha$ emission line (including the width), but also for all diagnostic lines at the same time. We include a similar requirement on the line width of both H$\alpha$ and [N ] as the single-Gaussian selection mode of Type-1 could have been the root of the observed differences, and by chance could result in the inclusion of non-AGN with poorly resolved H$\alpha$ and [N ]6585. However, increasing S/N ratio did not lead to $O_{colour} \sim 0$, and hence the emission line diagnostics are unlikely to cause any significant bias in our results.

### 1.7. Possible biases

At this point, one should question whether the different environments of Type-1 and Type-2 AGN might be caused by various kinds of selection effects.

### 1.7.1. Spectroscopic targeting

We first investigate whether the different targeting criteria in SDSS could bias our two parent AGN samples. As the Type-1 AGN sample may be dominated by specClass=3 objects ('Quasars') and the Type-2 AGN sample by specClass=2 objects ('Galaxies'), a bias in the selected neighbours may result. We therefore, for the two specClass categories separately, do the same colour-distance plots as in Figure 2 (Letter). We see again the same difference in colour of the the AGN neighbours. We therefore conclude that the targeting criteria is not the cause of the differing neighbour populations.

### 1.7.2. Mass-to-luminosity biases

One potential bias that could influence the selection of neighbours is if our Type-1 and Type-2 AGN of the same luminosity had different stellar masses of their host galaxies. The correlation between galaxy mass and environment could then be reflected in some of our results shown in the Letter, e.g. Fig 1 & 2.

There are several different ways in which this bias could be tested. We have already tested and excluded the possibility that any M/L bias between our samples– independent of the origin– could cause the colour differences between the two neighbour classes (Hypothetical Luminosity Test, see section 1.6). Here, one can compare two classes of objects for which biases could exist in the bolometric masses and luminosities. Since the test only attempted to match the neighbours of the two classes of objects, it stayed independent of the sources of contribution to the total luminosity of each object. Most luminosity biases (e.g. AGN continuum contribution in Type-1 AGN) can be tested with this approach.

However, as previously mentioned (see section 1.6), this method is new and relies on our Type-1 AGN not being luminous enough to outshine their host galaxy. While we are at low redshift where the risk of such occurrences is greatly reduced, this could still in some cases result in us comparing the luminosity of the Type-2 AGN host to the luminosity from the Type-1 AGN's nuclear region. Therefore, we perform additional tests with alternative methods and compare the conclusions.

One way to obtain the AGN hosts stellar masses would be to first deconvolve the power-law component from the spectra and then decompose the spectra with the help of generic AGN templates. However, this method has many pitfalls. While stellar synthesis codes have been very helpful in understanding the physical nature of the objects, there are large uncertainties in what is the the real AGN continuum contribution to the spectra (especially for Type-2 AGN) and how to disentangle the stellar and AGN component when estimating stellar masses of the AGN host (especially of Type-1 AGN). These pitfalls can cause the masses of the AGN in the two samples to be gravely biased relatively to each other. Therefore, we must match the masses by a method that is less dependent on prior assumptions about the object's physical nature – especially since our primary results suggest that the Type-1 and Type-2 AGN are very different type of objects and thus cannot motivate the underlying assumption that both are the same objects in the spectral decomposition procedure.

Therefore, we have adopted an alternative approach. AGN unification predicts that if the intrinsic AGN activity were the same for a Type-1 and Type-2 AGN, all the properties outside the obscuring material also inevitably must be the same. The $L$[O ] that arises in the narrow-line region outside the torus, is therefore believed to be one of the most accurate isotropic tracers of AGN activity (14). This means, that if we select two samples of Type-1 and Type-2 AGN purely based on matching each Type-1 to a Type-2 with the $L$[O ] and redshift, these *Type − 1 and Type − 2 AGN must also have the same stellar masses and intrinsic properties for the AGN unification to be true*. Hence, assuming AGN unification we can match the stellar masses by our AGN based on $L$[O ]. If AGN unification holds, the matched samples should represent galaxies of the same host stellar masses and therefore also must have the same neighbours (and the same amount of them).

We reselect our photometric neighbour samples based on $L$([O ]) *and* redshift of Type-1 and Type-2 AGN. For each Type-1 AGN we select the Type-2 AGN having the closest redshift and $L$([O ]). Only this fine matching and reselection where the stellar masses are predicted now to be closely matched, can give any conclusive arguments for or against the AGN unification Model. This means, we start off with exactly the same number of Type-1 and Type-2 AGN. The redshift and $L$([O ]) distributions are shown in Fig 5.

We retreive the neighbours for the newly reselected samples, and find that the number of pairs is 8503 for Type-1 AGN-galaxy pairs, while the corresponding number for Type-2 AGN-galaxy pairs is 7385 – a significantly lower number. We plot the colours of the neighbours having same absolute magnitude cut $M_r < −21.2$ as in Figure 2 (Letter). We again find that they are different, see Fig 6.

Could this difference between the neighbours be caused by a bias in $L$([O ])-matching arising from host plane dust extinction? (See Section 1.7.7 of SUPP. INF.) Our final and most conclusive test is to use only AGN with face-on spiral host galaxies matched in $L$([O ]) *and* redshift. First, we select only AGN that have spiral hosts according to Galaxy Zoo classification from our photometric neighbour samples. Then we visually classify each one of them as either "face-on" or "edge-on". We select then only those Type-1 and Type-2 spiral host AGN that we classify as "face-on", which corresponds to 86 Type-1 AGN and 1501 spiral and face-on Type-2 AGN. Now, we do the same matching based on $L$([O ]) and redshift as described above, and afterwards retrieve the neighbours. While the number of AGN hosts is small, we have many photometric neighbours for each object. We again see that the colours of the neighbours are different between Type-1 and Type-2 AGN, a difference that crystallizes in a plot very similar to Fig. 2 (Letter) or Fig 6 when applying the same absolute magnitude cut $M_r < −21.2$ on only the neighbour galaxies.

This proof-by-contradiction demonstrates that Type-1 and Type-2 AGN, with the same nuclear activity measured by $L$([O ]), do not reside in similar host galaxies and large-scale environments. Due to the failure of AGN unification in this test, the difference between the neighbours in this particular plot cannot be interpreted as the selection by $L$([O ]) bringing in an unknown bias. Only the figures in the Letter can therefore be used as clues for a physical interpretation.

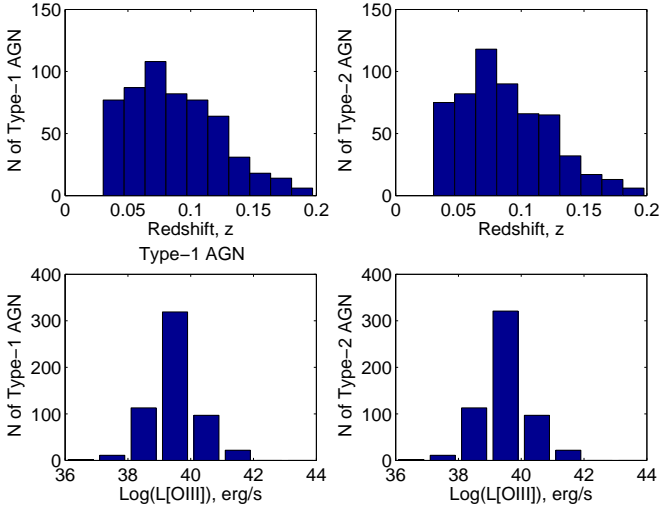

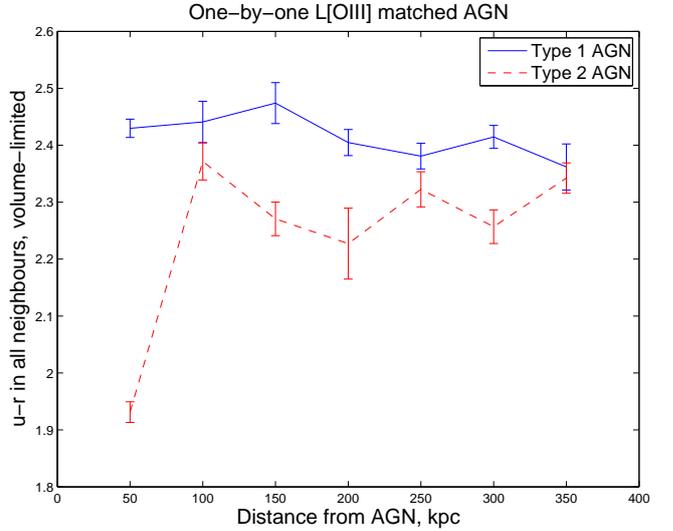

**Fig. 5.** Redshift and $L([O\;III])$-distributions of the 564 matched Type-1 and Type-2 AGN hosts in the pairs. Upper row shows the redshift distributions for the matched Type-1 (left) and Type-2 AGN (right). Lower row shows the $L([O\;III])$ distributions for the matched Type-1 (left) and Type-2 AGN (right).

**Fig. 6.** Mean $u - r$ colour of neighbour galaxies with $M_r < -21.2$ versus projected distance to the AGN for galaxies. The data is grouped in 50 kpc bins. Photometric redshift pairs with $|\Delta z| < 0.03$ are used in $L([O\;III])$ one-by-one matched samples of Type-1 and Type-2 AGN with the same redshift and $L([O\;III])$. The number of pairs in the plot is 1727 for the Type-1 AGN and 1289 for the Type-2 AGN. Standard errors for each bin are indicated by the $1\sigma$ error bars.

The Hypothetical Luminosity Test and the $L([O\;III])$ matching therefore agree in their conclusions.

### 1.7.3. Some additional test of luminosity biases

The $L([O\;III]5007)$ distributions support that the Type-1 and Type-2 AGN are selected within the same range of bolometric luminosities. But the Type-1 AGN will have a non-stellar continuum contribution in their spectra given that the contribution to the luminosity from the continuum is thought to be an orientation-dependent effect. At the same time, we argue that removing the continuum from the spectra could bias our samples (see previous section).

This means that at a given [O III] flux, the Type-1 AGN will always be a bit stronger than the Type-2 AGN. We investigate the average absolute magnitude for different ranges in $F([O\;III])$ in our volume-limited samples and see that indeed, the Type-1 AGN are somewhat brighter than the Type-2 AGN with the same [O III] flux. We wonder whether this could not cause a selection effect in favour of easier detection of Type-1 AGN.

We do some additional, rougher tests and compare the neighbours for AGN within overlapping ranges, this time *only* $L([O\;III])$ or $F([O\;III])$, but not matched in the redshift space.

Figure 3 shows the $L([O\;III])$ distribution of the Type-1 AGN and Type-2 AGN parent samples. We note that the biggest overlap of the luminosity occurs in the fifth bin, where the luminosity ranges between $10^{39} erg/s < L([O\;III]) < 10^{40} erg/s$. This narrow range in a volume-limited subsample allows us to investigate the colour-distance relationship the AGN with overlapping $L([O\;III])$. We observe that the conclusions do not change for these samples.

We also do the same investigation by selecting overlapping subsamples by selecting only those AGN in with different ranges of the $L([O\;III])$ and $F([O\;III])$. For the $F([O\;III])$ we used subsamples with AGN with $F([O\;III])$ in the ranges of 0–15 Å, 15–30 Å, 30–50 Å and more. Independent of how we select subsamples in [O III], the systematic differences between Type-1 and Type-2 AGN neighbours stay.

We test this by investigating the colour-distance relations of the AGN neighbours that are displaced in $L([O\;III])$. In the first investigation, Type-1 AGN have the $10^{38} erg/s < L[O\;III] < 10^{39} erg/s$ and Type-2 AGN have $10^{39} erg/s < L[O\;III] < 10^{40} erg/s$. The second analysis was for Type-1 AGN with $10^{39} erg/s < L[O\;III] < 10^{40} erg/s$ and Type-2 AGN with $10^{38} erg/s < L[O\;III] < 10^{39} erg/s$. In both cases, Type-1 AGN neighbours were significantly redder than the Type-2 AGN neighbours.

Finally, we repeat the test in the previous section where we used one-by-one $L([O\;III])$ matched spiral, face-on AGN that also have $L([O\;III]) > 10^{39.8} erg/s$. Here, the continuum contamination present in the Seyfert 1s is the smallest. The results again confirm that Type-1 AGN neighbours are significantly redder than the Type-2 AGN neighbours.

We conclude the bias from the stronger non-stellar continuum in Type-1 AGN is not dominant enough to influence the results to the observed extent. The results from the Hypothetical Luminosity Test (section 1.6) support this claim.

### 1.7.4. Weak emission lines

The most important biases might come from the sample selection itself and the use of weak emission lines in the classification.

While we used quite convential methods for creating the samples with the help of the Kauffmann (2003) criterion for the Type-2 AGN with $\sigma(H\alpha) < 10$ Å , the criterion used for the Type-1 AGN was single-Gaussian emission-line width $\sigma(H\alpha) > 10$ Å and thus depends on a single emission line measurement. The difficulties of resolving $H\alpha6565$ and [N ]6585 lines could cause a bias and one should possibly rather use multi-Gaussian fits for selecting broad-line AGN. Our approach to the problem was to examine if increasing the S/N ratio of the emission line fluxes and widths for both the $H\alpha6565$ and [N ]6585 emission lines would change the differences between the two samples in any way. If the choice of single-Gaussian line fit is the cause of selecting false broad-line AGN, increased demands on the S/N in both emission line widths and fluxes should decrease the number of potential false broad-line AGN. However, no changes were observed by introducing this criterion. Also, at earlier stages of the analysis we used $\sigma(H\alpha)=15$ Å as a limit, which yielded the same results.

We can also test the effect of the weak lines by simply increasing the minimum (lower) cut on the equivalent widths of the $H\alpha$ and $H\beta$ emission lines of the AGN and see how this influences the results. We try this both with the large samples as well as with the matched subsample with AGN within $10^{39} erg/s < L([O\ ]) < 10^{40} erg/s$.

If we increase the minimum $EW(H\alpha)$ we find that the neighbour colours at projected distances d < 50 kpc become more similar for Type-1 and Type-2 AGN neighbours. This was tried using lower cuts in $EW(H\alpha)$=5, 10, 15, 20, 25 and 30 Å. As this only is observed in the closest bin but not gradually for the Type-1 AGN neighbours, light contamination could be the reason. However at projected distances 50 < d < 350 kpc the Type-1 AGN neighbour colours stay significantly redder than the Type-2 AGN neighbour colours independent of the minimum $EW(H\alpha)$ used.

We also investigated the fraction of spiral hosts if varying the minimum $EW(H\alpha)$, and saw that it behaves in the same way in the presence/absence of a neighbour as it did in Fig. 4 (Letter).

We tried doing the same if varying the $H\beta$. However, with inreasing value of $EW(H\beta)$ the number of objects falls dramatically. If using minimum $EW(H\beta)=8$ Å and $EW(H\alpha)=30$ Å and make an isotropic selection with $10^{39} erg/s < L([O\ ]) < 10^{40} erg/s$ bin, we find that only 66 Type-1 AGN galaxy pairs and 79 Type-2 AGN galaxy pairs are left. Their average colours where for Type-1 AGN neighbours: $u − r \sim 2.4157 \pm 0.0949$ and for Type-2 AGN neighbours $u−r \sim 1.8000 \pm 0.0511$. If we again wish to avoid light contamination effects and only calculate the average colours for neighbours within projected distances of 50 < d < 350 kpc we have only 40 Type-1 AGN-galaxy pairs and 29 Type-2 AGN-galaxy pairs to use. Their colour is $u − r \sim 2.5283 \pm 0.0270$ and Type-2 AGN neighbours $u − r \sim 2.3344 \pm 0.0698$. This is consistent with our previous results that Type-1 AGN neighbours are redder on average than the Type-2 AGN neighbours.

Our conclusion is that our selected minimum equivalent widths on $H\alpha$ and $H\beta$ are not responsible for the differences in neighbours and that the systematic differences between the samples stay independent of the minimum emission line strength.

In a future publication, we will do the same analysis using more sophisticated broad-line samples (31; 12) and narrow-line samples (32; 33) as well as using the [O ] emission line for classification. The advantage of using these is that they also take into account contributions from stellar absorption and internal extinction from the host galaxy when constructing the samples, the neglect of which can be another bias in our sample selection.

The sample ratios of Type-1 to Type-2 AGN in our volume-limited cuts are similar to those found by several the other groups (31; 13), 1:2, which lends support to the robustness of our sample selection methods.

### 1.7.5. Composite objects

Many of our AGN are also having a strong star-formation and could be regarded as composite objects. It is difficult to remove the composite objects without introducing new biases, but we wonder whether there could exist a possible bias in favor of them with Type-2 as opposed to Type-1. Such a bias– if in favor of Type-2– could put our Type-2 AGN preferentially in star-forming environments and thus easier explain the difference in the environments in Figure 1 and 2. However, it still would fail to give an explanation to the morphological behavior in Figure 4, without implying a shorter time-scale for what we observe being the narrow-line population.

To investigate the potential effect of composite objects in our sample, we construct a new sample of Type-2 AGN that have the composite objects removed with criteria from Kewley et al (2006), but on the other hand include a number of LINERs. We find that the results are consistent with our previous ones– Type-1 AGN reside in significantly redder environments than Type-2 AGN. Therefore, no bias in favour of composite objects among the Type-2 AGN as opposed to Type-1, is sufficient to explain this difference.

### 1.7.6. Sky-covering factor, background galaxies and light contamination

Also the sky-covering factor could be a cause of seeing different neighbours around Type-1 and Type-2 AGN. However, the large differences of morphologies of the AGN hosts rules out this pos-sibility. Redshift and luminosity biases can be equally ruled out as they would not produce a correlation in the ratio Type1/Type2 neighbours around Type-2 as in Figure 3 (Letter).

Another bias we considered is whether the light from a Type-2 AGN would contaminate the closest neighbours (the bin at the shortest projected separation) and therefore create a false trend in the colours. However, we examine the luminosity distributions of AGN hosts and note that the Type-2 AGN are not any more luminous than the Type-1 AGN, thus light contamination can not be the cause of the differences in the closest bin. To check if background galaxies might be the cause, we plots the spectroscopic pairs in three different redshift cuts $|\Delta z| < 0.001$, $|\Delta z| < 0.006$ and $|\Delta z| < 0.012$, which all yield the same results.

### 1.7.7. Effect of dust

An important effect that might play a substantial role in the selection of the samples is the dust content the host galaxies or the dust content in the large-scale environment. If we for instance imagine the light from a BLR on its way to us would pass through the dusty plane of a spiral galaxy, this could make the Type-1 easily appear as a Type-2 AGN. In such way, we could ask whether not a large part of the Type-2 AGN are actually observed as Type-2 AGN only because they reside in dustier host galaxies? Qualitatively, also the presence of a companion could act in a similar way by obscuring the BLR emission from our way.

This would agree with some of our observations: the fact that only a very few Type-2 AGN are observed in elliptical host galaxies. Also, it could in some way explain why the ratio of Type-1/Type-2 AGN neighbours decreases with at close projected separations from the Type-2 AGN; the obscuration of Type-1 AGN nuclei would transform it into a Type-2 AGN in Fig.3 (Letter). However, the effect of dust obscuration from the host galaxy itself still fails to give a satisfactory explanation to the inconsistent morphological behavior in Fig.4 (Letter) as well as the colour-distance plots in Fig.1 (Letter) and Fig.2 (Letter). This suggests that there must exist more fundamental differences between our Type-1 and Type-2 AGN neighbour samples than sampling effects due to dust obscuration.

In section 1.7.2 we tested whether the inclination of spiral host galaxies change our results or not for $L([O\ ])$-matched samples, since extinction in the dust host plane could make some Type-1s appear as Type-2s. We there that even if we select only face-on spiral galaxies and look at their neighbours, the colour of the neighbours consistently stays the same.

### 1.7.8. Mixing radio-loud and radio-quiet

In this study we have not separated radio-quiet from radio-loud objects. Such a study would require much higher-precision radio measurements on all the AGN hosts than available in the catalogues today. A bias from including more radio-loud objects into one of the samples could in principle influence our results and conclusions. In the future, we hope to do this with a better categorization of a large number of objects.

However, since we do studies separately for spiral AGN hosts that extremely rarely are radio-loud and still see the same differences in the neighbour colour, we know that mixing of radio-loud/radio-quiet objects is not a sufficient bias to lead to our conclusions.

### 1.7.9. Covering factor

From the Realistic unification we know of the difference in covering factors in the tori of different AGN might cause a luminosity bias in the selection of AGN, and therefore could cause a different clustering of neighbours. We approached the problem in two ways: by testing with matched $L([O\ ])$ subsamples and with the hypothetical luminosity test. Even if performing the test on host galaxies initially (by only investigating the colours of neighbours) suggests a luminosity displacement corresponding to $E_{dis} \sim 4$ mag in Type-2 AGN, this reason can be rejected. The morphology differences of AGN host galaxies between the two population stay different irrespective of whether we select Type-1 and Type-2 AGN with "colour-matched" neighbour populations. And if we perform the hypothetical luminosity test only for spiral hosts, there turns out to never exist a luminosity displacement that can yield the same neighbour colours.

However, one would need to redo this test using stellar masses of the host galaxies to provide further support for our statement.

### 1.8. What happens to the torus?

One could wonder, if there truly exists also a temporal relationship between Type-1 and Type-2 AGN, what happens to the dust structure during interactions?

For instance, the geometrically thick outflow model (34) has been very successful in predicting some AGN-features e.g. water masers connected to the torus complex as well as the disappearance of the torus at low luminosities $L < 10^{42}$ erg/s. If the inflow of gas would be moderate or not too strong, it is possible that an interaction/merger would be more likely to increase the accretion rate of the accretion disk and thus also thicken the torus. This means, that interactions were the fueling is not dominant would increase the covering factor of the torus and the AGN's probability of being observed as Type-2 AGN. Such interactions, might also be insufficient to change the overall morphology of the host galaxy. But within the same geometrically thick outflow model, what would happen to the BLR/torus complex if the sudden inflow of gas to the accretion disk would be exceptionally large and give rise to a very strong accretion rate? One could imagine, that such an extreme inflow and increased accretion rate in the accretion disk could lead to sublimation (and possibly ionization) of the dust in the torus. An inflow of gas so large would then not only transform the Type-2 AGN into a Type-1

AGN, but also be sufficient to transform the previous spiral into an elliptical galaxy.

This scenario be consistent with our observations of how the fraction of AGN residing in spiral hosts differs between Type-1 and Type-2 AGN. It highlights the importance of developing the theory for an evolutionary transformation between the two AGN types.